# Long-lived nonequilibrium spin configuration induced by dual optical excitation

Sergii Parchenko[1*], Arata Tsukamoto[2], Andreas Scherz[1]

[1]*European XFEL, Holzkoppel 4, 22869 Schenefeld, Germany*

[2]*College of Science and Technology, Nihon University, 7-24-1 Funabashi, Chiba 274–8501, Japan*

*Corresponding author: sergii.parchenko@xfel.eu

**Abstract.** The ability to create and control nonequilibrium spin states is central to understanding ultrafast magnetism and to developing optical routes for manipulating magnetic materials beyond conventional thermal pathways. In metallic systems, photoexcited electronic distributions rapidly evolve toward quasi-thermal states through efficient electron–electron and electron–phonon scattering, making long-lived modification of spin-dependent interactions difficult to achieve. Here, we investigate coherent magnetization precession in an amorphous GdFeCo thin film driven by simultaneous excitation with two femtosecond laser pulses. Using a symmetric dual-pump geometry that continuously evolves from single-pulse to spatially periodic excitation, we separate effects associated with the excitation pattern from those arising from simultaneous dual-pump excitation. We identify a narrow excitation window in which the lifetime of the ferromagnetic-resonance–like mode increases strongly while its frequency exhibits a small but reproducible increase. Time-window analysis shows that the enhanced lifetime persists for several hundred picoseconds, substantially longer than transient magnetostatic effects associated with the periodic excitation pattern. The combined frequency and decay time changes cannot be explained by spatially structured heating alone and instead point to a long-lived nonequilibrium spin-dependent electronic configuration. Within a phenomenological Elliott–Yafet framework, the simultaneous increase in FMR decay time and frequency is consistent with a transient reduction of the effective spin-mixing parameter governing spin-flip scattering. Our results demonstrate that the excitation pathway, rather than excitation energy alone, can serve as a control parameter for spin relaxation and provide access to nonequilibrium spin states that are not readily reached by conventional single-pulse excitation.

## I. Introduction

Ultrafast optical excitation provides a unique means of driving magnetic materials far from equilibrium and manipulating spin dynamics on their intrinsic femtosecond timescales [1,2]. During the past three decades, this approach has enabled the observation of phenomena ranging from ultrafast demagnetization [3,4,5] and transient modification of magnetic anisotropy [6,7] to coherent excitation of collective spin modes [8,9,10,11,12,13]. Besides their potential for ultrafast spintronic technologies [13,15,16,17,18], these experiments address a fundamental problem of condensed-matter physics: how angular momentum is redistributed among electronic, spin, orbital, and lattice degrees of freedom following an ultrafast perturbation. A central challenge in this field is the creation and control of nonequilibrium spin-dependent electronic states that persist beyond the initial excitation [19,20,21,22]. In metallic systems, photoexcited carrier distributions typically evolve on extremely short timescales owing to efficient electron–electron and electron–phonon scattering and the large density of electronic states around the Fermi level [23]. Within only a few picoseconds, the electronic system approaches a quasi-thermal distribution, after which the subsequent magnetic dynamics are governed predominantly by heat transport and lattice cooling. Consequently, most optically induced electronic configurations are inherently short-lived, making it difficult to manipulate microscopic spin-dependent interactions independently of conventional thermal effects.

Recent experiments suggest that simultaneous excitation with two coherent femtosecond laser pulses may provide a fundamentally different route for controlling nonequilibrium spin states [24,25,26,27]. In such dual-pump excitation, two optical fields interact with the material simultaneously while forming a spatially periodic excitation profile. Prior studies have reported a pronounced slowing of magnetic relaxation compared with single-pulse excitation at comparable total absorbed fluence, including both the recovery following ultrafast demagnetization [25] and the damping of coherent spin dynamics [24]. These observations raise the intriguing possibility that the excitation pathway itself, rather than the deposited energy alone, transiently accesses spin-dependent electronic configurations that are not populated under conventional single-pulse excitation. The interpretation of these experiments, however, remains challenging. Dual-pump excitation inevitably produces a periodic distribution of deposited energy, introducing spatial variations of temperature, magnetic anisotropy, and demagnetization fields that can themselves modify the observed dynamics [28,29,30,31]. Distinguishing these geometric and thermal effects from an intrinsic modification of the electronic spin state is therefore nontrivial. At the same time, a microscopic theoretical description of dual-pump excitation remains beyond present computational capabilities. While first-principles approaches such as time-dependent density functional theory have substantially advanced the understanding of ultrafast spin dynamics, they are currently restricted to atomistic length scales and sub-picosecond dynamics. Experimentally relevant conditions, including micrometer-scale excitation patterns, collective spin dynamics extending over hundreds of picoseconds, and the simultaneous treatment of nonequilibrium electronic, spin, and lattice degrees of freedom, require a multiscale description that is not yet available [32]. Consequently, phenomenological approaches remain essential for identifying the microscopic mechanisms underlying complex nonequilibrium magnetic phenomena.

In this work, we address this question using coherent magnetization precession in amorphous GdFeCo thin film as a sensitive probe of nonequilibrium spin dynamics. A symmetric dual-pump geometry is employed in which the fluence of one excitation pulse is continuously varied while the second pulse is kept fixed, allowing the periodic excitation pattern to evolve smoothly from the single-pump limit to a high-contrast interference pattern. This experimental strategy enables the effects arising from spatially structured excitation to be distinguished from an additional contribution associated with dual optical excitation itself. We identify a narrow excitation window in which the damping of the FMR-like mode is strongly reduced while the precession frequency exhibits a small but reproducible increase. Time-window analysis further demonstrates that the damping reduction persists for several hundred picoseconds, substantially exceeding the lifetime of transient magnetostatic effects associated with the periodic excitation pattern. We discuss these observations within a phenomenological framework based on Elliott–Yafet spin-flip scattering. The combined increase of the FMR decay time and the accompanying frequency shift are consistent with a transient reduction of the effective spin-mixing parameter governing spin-flip probability, suggesting that dual optical excitation drives the system into a long-lived nonequilibrium spin-dependent electronic configuration inaccessible through conventional single-pulse excitation. More broadly, our results establish excitation pathway, as opposed to excitation energy alone, as a controllable parameter for engineering spin relaxation in metallic magnetic systems and provide a framework for exploring optically generated nonequilibrium spin states beyond conventional thermal descriptions.

## II. Experimental methods

The experimental geometry is schematically shown in Fig. 3a. Ultrafast magnetization dynamics were measured using a pump–probe magneto-optical setup based on an ActiveFiber laser delivering 1030 nm pulses of ~300 fs duration at 50 kHz. A fraction of the beam was frequency-doubled to 515 nm for time-resolved probing. Dual-pump excitation was realized by splitting the pump into two arms (Pump 1 and Pump 2) recombined on the sample at symmetric angles of +30° and −30° relative to the surface normal. This geometry generates a high-contrast periodic excitation pattern while avoiding the asymmetry of previous configurations. Pump and probe spot radii were ~250 μm and ~100 μm, respectively, ensuring homogeneous probing. Direction of linear polarization of the pumps was controlled with a half-wave plate, and the probe was also linearly polarized. Pump modulation at 500 Hz enabled lock-in detection of small magneto-optical signals. Time-resolved magnetization dynamics were accessed via magneto-optical Kerr effect. The reflected probe was split into orthogonal polarizations using a Wollaston prism and detected with a balanced photodiode, providing the transient Kerr rotation and thus the out-of-plane magnetization component. A 520 ± 40 nm bandpass filter suppressed scattered pump light. The sample under investigations is a 20 nm amorphous $Gd_{23}(FeCo)_{77}$ thin film on Si with a 3 nm $Si_3N_4$ capping layer. For the studied composition, the film exhibits perpendicular magnetic anisotropy, with the magnetization compensation temperature $T_M$ and angular momentum compensation $T_A$ below room temperature [33,34]. External magnetic field was applied at an angle about 60 deg from sample normal. All experiments were carried out in room temperature.

## III. Results

### A. Single-pump FMR-like dynamics in GdFeCo

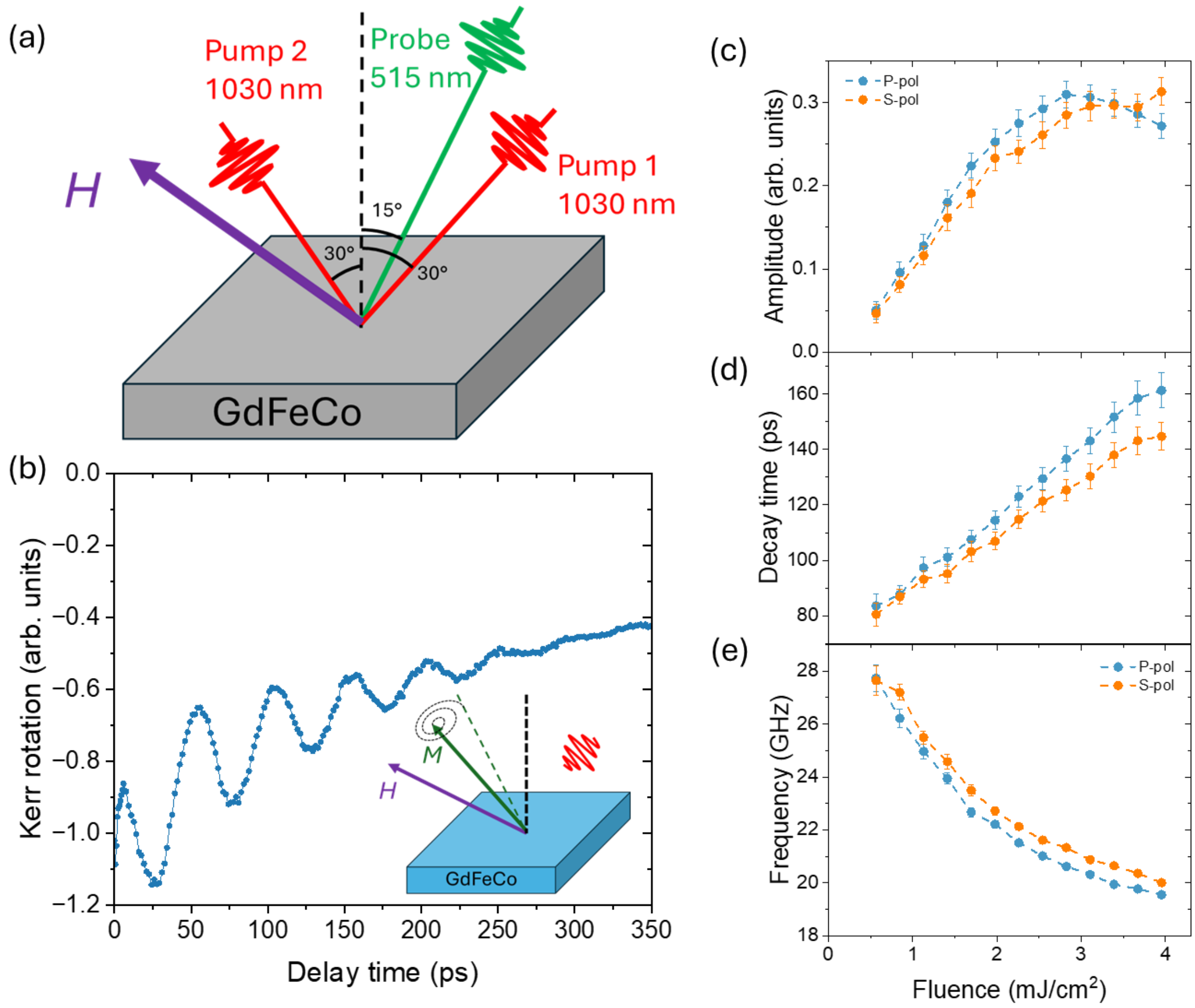

***Figure 1. | Single-pump FMR-like precessional dynamics in GdFeCo.*** *(a) Schematic of the experimental configuration with dual ultrafast pump pulses and time-resolved Kerr rotation probe. (b) Time-resolved Kerr rotation signal showing coherent oscillatory precession on top of a slowly varying background after single-pump excitation. (c–e) Fluence dependence of precession parameters extracted from damped sinusoidal fits: (c) precession amplitude, (d) decay time, and (e) frequency. The inset to panel (b) explains the mechanism of magnetization precession excitation after optical excitation.*

We first establish the single-pump response of the GdFeCo film. Optical excitation induces an ultrafast reduction of the magnetization magnitude together with a transient modification of the effective magnetic anisotropy. As a result, the equilibrium orientation of the magnetization is abruptly shifted, and the magnetization starts to precess around the new effective field direction following Landau-Lifshitz formalism (see inset to Fig. 1b). This excitation mechanism is well established for ferromagnetic and ferrimagnetic thin films. A representative time-resolved Kerr rotation trace measured under single-pump excitation is shown in Fig. 1b. The signal exhibits a pronounced oscillatory component, characteristic of coherent precession, on top of a slowly varying background associated with the gradual recovery of the magnetization to its equilibrium orientation. To quantitatively analyze the dynamics, the data were fitted using a damped sinusoidal function combined with an exponential background. From these fits, we extract the fluence dependence of the precession amplitude, decay time, and frequency, as summarized in Figs. 1c–e. The precession amplitude increases with increasing pump fluence and shows a tendency toward saturation at higher excitation densities. The precession frequency decreases monotonically with fluence, reflecting the reduction of the magnetization magnitude and the associated softening of the effective internal magnetic fields as the sample is heated. At the same time, the decay time of the precession increases with fluence, indicating a gradual reduction of the effective damping with increasing excitation density. All observed trends are consistent with previously reported single-pump studies of optically induced precession in GdFeCo and related ferrimagnetic systems.

### B. Dual-pump–induced modification of FMR dynamics in GdFeCo

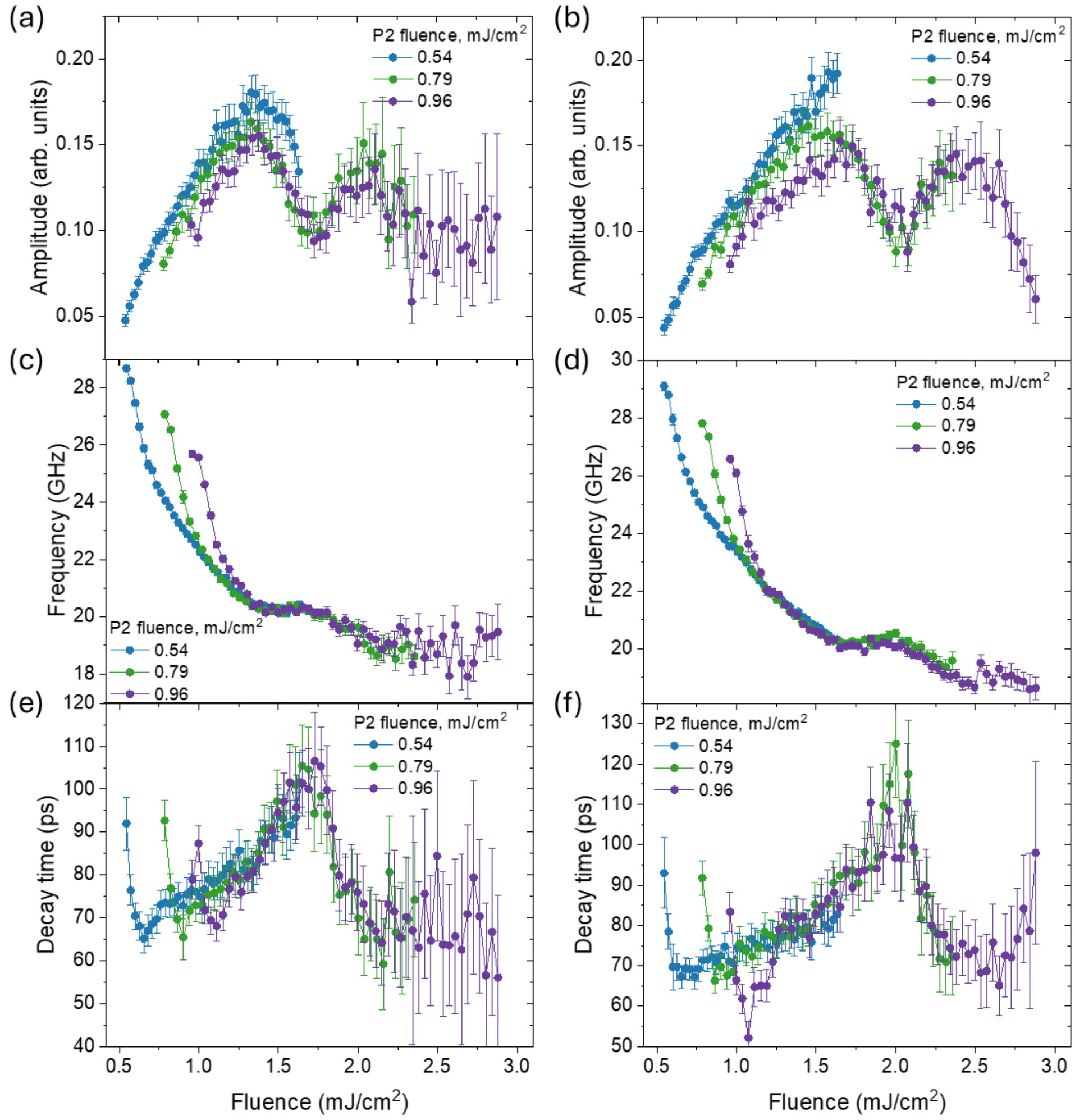

***Figure 2. Dual-pump modification of precessional dynamics in GdFeCo.*** *(a,b) Precession amplitude as a function of combined fluence for P–P (a) and S–S (b) polarizations at three fixed P2 fluences (0.54, 0.79, and 0.96 mJ/cm$^2$). (c,d) Corresponding precession frequencies. (e,f) Decay times of the FMR-like mode. A pronounced modification of all precessional parameters occurs within a well-defined total fluence window, including amplitude suppression, slight frequency increase, and dramatic elongation of decay time. These changes cannot be explained by heating or periodic excitation pattern alone, indicating a dual-pump–induced modification of spin relaxation.*

We now turn to the dual-pump excitation regime, which constitutes the central result of this work. The experimental protocol is the following: we record magnetization precession traces when the fluence of Pump 2 (P2) is fixed, while the fluence of Pump 1 (P1) is gradually increased from zero. Figure 2 presents results for three representative P2 fluences, 0.54 mJ/cm$^2$ (blue), 0.79 mJ/cm$^2$ (green), and 0.96 mJ/cm$^2$ (purple). Measurements were performed for both P–P and S–S polarization configurations of the two pump beams. From the time-resolved Kerr rotation traces, we extract the precession amplitude, frequency, and decay time as a function of the total incident fluence. The corresponding results are summarized in Fig. 2: panels (a,b) show the precession amplitude for P–P and S–S excitation, panels (c,d) the precession frequency, and panels (e,f) the decay time. At low P1 fluence, when P1 is first added to P2, the observed trends largely follow the single-pump behavior. The precession amplitude increases with total fluence, while the precession frequency decreases due to increased heating and the associated reduction of the effective internal magnetic fields. In this initial regime, however, a modest reduction of the decay time is observed for several data points. This decrease is related to developing of a periodic excitation pattern and is attributed to enhanced dephasing of the uniform precessional mode caused by spatial modulation of the demagnetization field. As regions with different excitation densities precess with different initial deflections, the resulting inhomogeneous stray-field landscape leads to increased dephasing and thus a reduced apparent decay time. This effect is expected from the formation of a periodic excitation profile alone and does not require any additional microscopic interaction between the two pump pulses. Importantly, it establishes a well-defined baseline for identifying nontrivial dual-pump effects.

Upon further increasing the P1 fluence, the system enters a distinct and well-defined fluence window in which the magnetization dynamics deviates qualitatively from the expected behavior. Within a combined fluence range of approximately 1.3–2.0 mJ/cm$^2$ for P–P excitation and 1.6–2.3 mJ/cm$^2$ for S–S excitation, all three precessional parameters are simultaneously modified. In this regime, the precession amplitude is strongly suppressed, the precession frequency exhibits a small but systematic increase, and, most notably, the decay time increases dramatically. This behavior is reproducible for all three fixed values of the P2 fluence and occurs at essentially the same total fluence range in each case. A key observation is that, just below and just above this fluence window, the amplitude, frequency, and decay time collapse onto the same values for all P2 fluences, indicating that outside this window the dynamics is governed solely by the total excitation density. In contrast, the periodic excitation pattern increases monotonically with increasing P1 fluence over the entire range. The confinement of the anomalous behavior to a narrow fluence window therefore demonstrates that it cannot be explained by heating or magnetostatic dephasing effects alone. We conclude that dual-pump excitation introduces an additional mechanism that counteracts the dephasing associated with the periodic excitation pattern and leads to a pronounced prolongation of the precessional lifetime within a specific excitation window. The sharp fluence selectivity and the simultaneous modification of amplitude, frequency, and decay time point to a transient alteration of the underlying spin dynamics that is not accessible under single-pump excitation.

### C. Temporal extent of the dual-pump effect

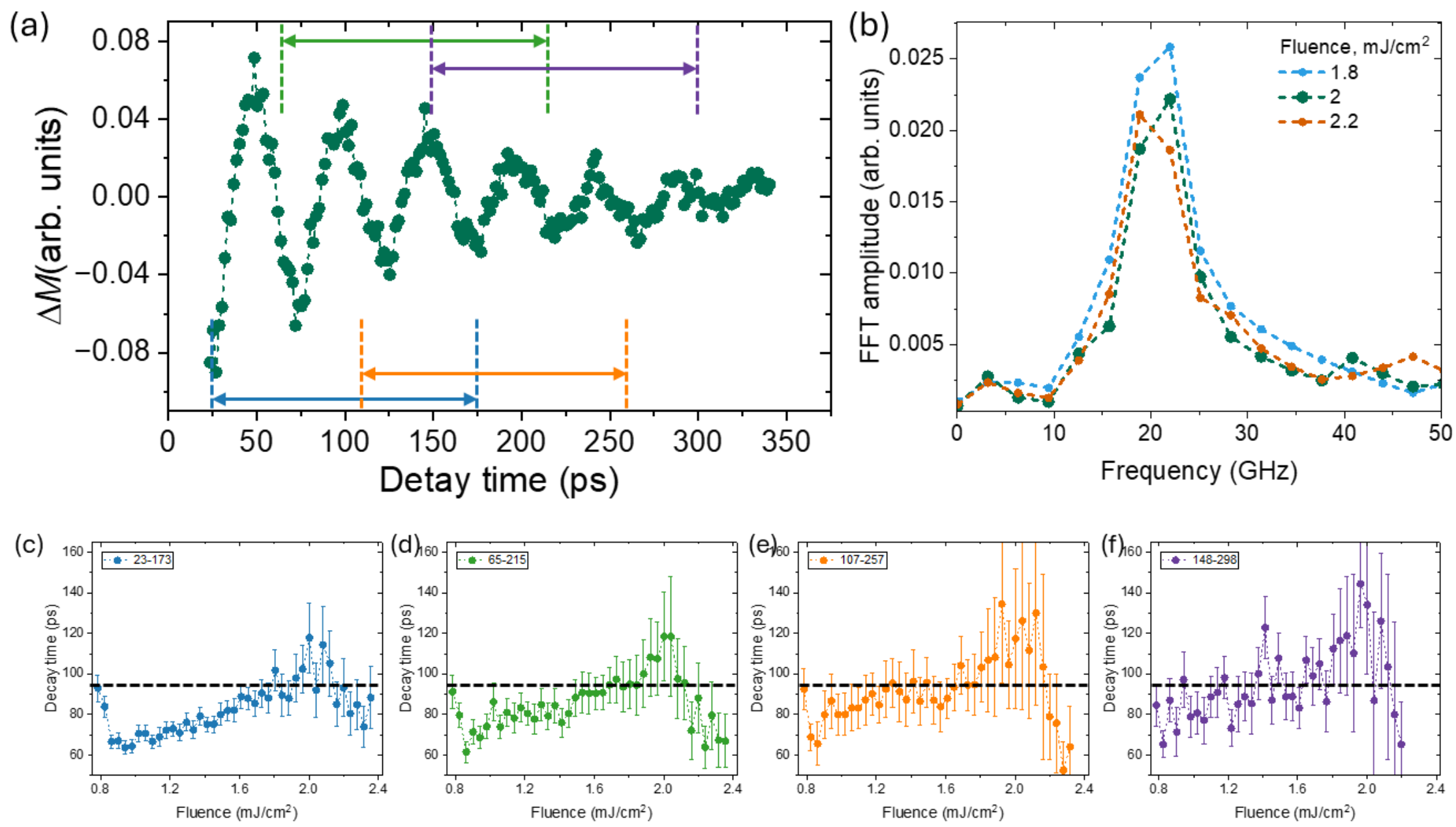

***Figure 3. Temporal persistence of the dual-pump effect in GdFeCo.*** *(a) Segment of the Kerr rotation oscillation (S–S polarization,* $F_{P2}$ *= 0.79 mJ/cm*$^2$*,* $F_{total}$ *= 2 mJ/cm*$^2$*) after subtraction of the slowly varying background, highlighting the coherent precession. Colored arrows indicate sliding time windows used for fitting. (c–f) Extracted decay times from damped sinusoidal fits within successive 150-ps windows. While initial reduction of decay time due to periodic excitation rapidly diminishes, the dual-pump–induced enhancement persists across the entire 300-ps window. Fourier analysis confirms a single dominant precession mode, indicating the long-lived modification arises from intrinsic changes in damping rather than mode superposition.*

We next examine the temporal extent of the dual-pump–induced modification of magnetization dynamics in GdFeCo. The analysis is summarized in Fig. 5. Figure 5a shows a representative segment of the magnetization dynamics for the S–S polarization configuration, where the dual-pump effect is most pronounced. The data correspond to a fixed P2 fluence of 0.79 mJ/cm$^2$ and a combined fluence of 2 mJ/cm$^2$, i.e. within the fluence window where the enhancement of the decay time is maximal. To isolate the precessional dynamics, the slowly varying exponential background has been subtracted, leaving only the oscillatory component. To track the temporal evolution of the effect from dual optical excitation, we perform damped sinusoidal fits within sliding time windows of 150 ps duration, as illustrated by the colored arrows in Fig. 3a. Four representative fitting windows are analyzed, spanning the delay ranges 23–173 ps, 65–215 ps, 107–257 ps, and 148–298 ps. The decay times extracted from these fits are shown in Fig. 3c–f. In all four fitting windows, a clear enhancement of the decay time is observed within the same fluence range identified in Fig. 2. This remains the case despite the substantial overlap between the fitting windows. As the analysis window is shifted to later delays, the oscillation amplitude decreases and the associated uncertainties increase. Nevertheless, the enhancement of the decay time in the dual-pump fluence window remains clearly discernible. Importantly, within the experimental uncertainty, the magnitude of the decay-time enhancement does not show a systematic reduction when moving the fitting window to longer delays, indicating that the effect persists throughout the entire investigated time range. By contrast, the initial reduction of the decay time observed at low P1 fluence, attributed above to magnetostatic dephasing associated with the formation of a periodic excitation pattern, exhibits a distinctly different temporal behavior. As the fitting window is shifted to later delays, this reduction rapidly weakens and approaches the decay time measured for P2-only excitation, indicated by the dashed line in Fig. 3c–f. Already in the second fitting window the reduction is strongly suppressed, and for windows starting beyond approximately 100 ps it is nearly absent. This demonstrates that the influence of the periodic excitation pattern on the decay time is confined to early delay times below ~100 ps. The clearly different temporal evolution of these two contributions allows them to be disentangled. While magnetostatic dephasing associated with the periodic excitation pattern dominates the early-time response and decays rapidly, the dual-pump–induced enhancement of the decay time persists on much longer time scales, extending to at least 300 ps.

To further exclude the possibility that the increased decay time arises from mode beating or spectral broadening due to spatially nonuniform excitation, we analyze the Fourier spectra of the oscillatory signal over the full delay window for fluences just below the dual-pump regime, at the maximum of the effect, and just above it. In all cases, the FFT reveals a single dominant frequency peak with comparable linewidth. This confirms that the observed behavior does not originate from the superposition of multiple precessional modes or from spatial averaging effects. Taken together, these results demonstrate that dual-pump excitation induces a genuine and long-lived modification of the intrinsic damping of the FMR-like mode in GdFeCo. This effect is clearly distinct from transient magnetostatic dephasing associated with periodic excitation and persists on time scales of several hundred picoseconds.

## IV. Discussion:

In metallic magnetic systems, relaxation of coherent spin dynamics is fundamentally governed by the transfer of angular momentum from the electronic spin system to other degrees of freedom, most importantly the lattice. A widely used microscopic framework for describing this process is the Elliott–Yafet (EY) mechanism, where spin–orbit coupling provides the spin-mixing required for spin-flip scattering during otherwise spin-conserving momentum relaxation events [35,36,37,38,39,40,41,42]. Because of spin–orbit coupling, electronic eigenstates in a crystal are not pure spin states. A Bloch state with predominantly spin-up character can be written schematically as

$$|\psi_{k,\uparrow}\rangle = a_k|\uparrow\rangle + b_k|\downarrow\rangle,$$

where $|b_k|^2 \ll 1$,represents the spin-mixing parameter. The coefficient $b_k$ originates from the spin–orbit interaction,

$$H_{SO}=1/(2m^2c^2)(\nabla V(r)\times p)\cdot S,$$

which couples the electron spin to the orbital motion in the crystal potential. As a consequence, even scattering processes that do not directly act on the spin degree of freedom, such as electron–phonon or impurity scattering, acquire a finite probability to transfer angular momentum between spin states. Within the Elliott–Yafet framework, the spin-flip scattering rate can be expressed phenomenologically as

$$\tau_{sf}^{-1}\propto\langle|b|^2\rangle\tau_p^{-1},$$

where $\tau_p$ is the momentum relaxation time and $\langle|b|^2\rangle$ represents an average spin-mixing parameter over the electronic states involved in the nonequilibrium dynamics. Therefore, a reduction of the effective spin-mixing parameter leads to a lower spin-flip probability and a longer characteristic spin relaxation time.

The damping of coherent magnetization precession can be connected to the same microscopic processes [40,41,42]. In a macroscopic description, the decay of an FMR-like mode is characterized by the Gilbert damping parameter α,

$$\tau_{FMR}\approx 1/(\alpha\omega),$$

where ω is the precession frequency. Microscopically, the damping parameter describes the dissipation of coherent spin motion and is related to angular momentum transfer from the spin system to electronic and lattice reservoirs. In phenomenological EY-based descriptions, the damping is connected to the efficiency of spin-flip scattering and can be expressed as $\alpha\propto\langle|b|^2\rangle$. Therefore, a transient reduction of the effective spin-mixing parameter provides a natural explanation for a simultaneous increase of the precession decay time and suppression of spin relaxation.

The central observation of this work is that dual optical excitation modifies the damping of coherent magnetization precession beyond what is expected from increased excitation density or from the formation of a periodic excitation profile alone. Within the EY framework, the observed increase of the decay time implies that, under otherwise comparable conditions, the effective spin-mixing parameter is reduced:

$$\langle|b|^2\rangle_{dual}<\langle|b|^2\rangle_{single}$$

This relation should be interpreted as an effective property of the transient nonequilibrium electronic distribution rather than as a modification of the intrinsic spin–orbit interaction. Within this phenomenological picture, dual-pump excitation transiently reduces the effective spin mixing relevant for momentum-relaxing scattering events, thereby lowering the spin-flip probability. A natural consequence is an increase of the precessional lifetime, consistent with the experimentally observed reduction of FMR damping. Moreover, the simultaneous increase of the precession frequency within the same fluence window suggests that the dual-pump state cannot be described as a purely thermally excited magnetic state, for which heating alone would lead to a monotonic frequency reduction through magnetic softening. Previous temperature- and composition-dependent studies[33,34] of GdFeCo have shown that approaching the angular-momentum compensation condition leads to simultaneous increases in the FMR-like precession frequency and effective Gilbert damping, resulting in a shorter relaxation time. This behavior is qualitatively different from the present dual-pump regime, where the precession frequency increases while the lifetime is markedly prolonged. Thus, although ferrimagnetic compensation effects may contribute to the fluence dependence of the precessional parameters, the observed combination of frequency enhancement and prolonged lifetime is not consistent with a simple thermal evolution toward the angular-momentum compensation condition.

The microscopic origin of this modified spin-mixing state remains an open question. A plausible interpretation is that simultaneous excitation by two coherent optical fields with different propagation directions generates a nonequilibrium electronic distribution that differs qualitatively from that produced by a single pulse. Since optical transitions couple directly to electronic momentum states through dipole matrix elements, the excitation pathway may modify the relative population of states characterized by different spin-mixing strengths. In this scenario, the effective parameter $\langle |b|^2 \rangle$ changes because the statistical weighting of participating electronic states changes, rather than because the spin-mixing coefficient of every state is altered. The observation of the effect for both P–P and S–S polarization configurations indicates that this redistribution is not determined solely by the orientation of the optical electric field but more generally by the simultaneous action of two coherent excitation pathways. At present, this interpretation remains qualitative because our measurements probe only momentum-averaged observables. Direct verification will require future momentum-resolved experiments and corresponding theoretical developments.

Finally, the persistence of the reduced damping for several hundred picoseconds points to the formation of a metastable nonequilibrium electronic state. Electron–electron and electron–phonon equilibration in metals typically occur within only a few picoseconds, making delayed thermalization an unlikely explanation. Together with the narrow fluence selectivity of the observed effect, this long lifetime distinguishes the dual-pump state from conventional heating-induced dynamics and suggests that the excitation pathway itself can be used to stabilize nonequilibrium spin states and control magnetic relaxation beyond purely thermal limits.

## V. Conclusion

In summary, we have demonstrated that simultaneous excitation with two coherent optical pulses modifies the decay time of coherent magnetization precession in GdFeCo in a manner that cannot be explained by the combined effects of increased fluence or spatially periodic heating. By continuously evolving the excitation from the single-pump to the dual-pump regime in a symmetric excitation geometry we isolate a narrow fluence window in which the magnetic dynamics deviates qualitatively from conventional thermal behavior. The observed increase of decay time, accompanied by a small increase of the precession frequency and persisting for several hundred picoseconds, indicates that dual optical excitation transiently drives the system into a distinct nonequilibrium electronic configuration with modified spin-relaxation pathways. Within the Elliott–Yafet phenomenology, these observations are consistent with a transient reduction of the effective spin-mixing parameter governing spin-flip scattering. Although the microscopic origin of this modification remains unresolved, the results demonstrate that the excitation pathway itself constitutes an additional degree of freedom for engineering nonequilibrium electronic states. Rather than simply depositing energy into the material, coherent multi-field excitation appears capable of steering the electronic system toward configurations that are not accessible through conventional single-pulse excitation.

These findings suggest that dual optical excitation provides access to a previously unexplored regime of nonequilibrium magnetism, in which the electronic state, and consequently the magnetic interactions that emerge from it, can be transiently modified beyond the limits of quasi-thermal excitation. Since exchange interactions, magnetic anisotropy, damping, and other collective magnetic properties all originate from the underlying electronic structure, tailoring this transient electronic configuration offers a potential route toward manipulating multiple magnetic couplings simultaneously. More broadly, the present work establishes coherent multi-field excitation as a promising strategy for exploring and controlling electronic states that remain hidden under conventional excitation conditions, opening new directions for ultrafast spin dynamics and nonequilibrium condensed-matter physics.

**Acknowledgments.**

The authors thank Prof. Peter Oppeneer for helpful discussions.

**References**

1. A. Kirilyuk, A. V. Kimel, and T. Rasing, Ultrafast optical manipulation of magnetic order, Reviews of Modern Physics **82**, 2731 (2010).
2. A. Zong, B. R. Nebgen, S.-C. Lin, J. A. Spies, and M. Zuerch, Emerging ultrafast techniques for studying quantum materials, Nat Rev Mater **8**, 224 (2023).
3. E. Beaurepaire, J.-C. Merle, A. Daunois, and J.-Y. Bigot, Ultrafast Spin Dynamics in Ferromagnetic Nickel, Physical Review Letters **76**, 4250 (1996).
4. J. Hohlfeld, E. Matthias, R. Knorren, K.H. Bennemann, Nonequilibrium magnetization dynamics of nickel, Phys. Rev. Lett. **78**, 4861 (1997).
5. J.-Y. Bigot, M. Vomir, and E. Beaurepaire, Coherent ultrafast magnetism induced by femtosecond laser pulses, Nature Phys **5**, 515 (2009).
6. V. N. Kats et al., Ultrafast changes of magnetic anisotropy driven by laser-generated coherent and noncoherent phonons in metallic films, Phys. Rev. B **93**, 214422 (2016).
7. Y. W. Windsor et al., Deterministic control of an antiferromagnetic spin arrangement using ultrafast optical excitation, Commun Phys **3**, 139 (2020).
8. V. Wiechert, H. Wang, W. Legrand, P. Gambardella, D. Breitbach, P. Pirro, M. Lammel, A. Meo, G. Finocchio, and D. Bossini, On demand laser-induced frequency tuning of coherent magnons in a nanometer-thick magnet at room temperature, Nat Commun **17**, 145 (2026).
9. A. Mekonnen, M. Cormier, A. V. Kimel, A. Kirilyuk, A. Hrabec, L. Ranno, and Th. Rasing, Femtosecond Laser Excitation of Spin Resonances in Amorphous Ferrimagnetic $Gd_{1-x}Co_x$ Alloys, Phys. Rev. Lett. **107**, 117202 (2011).

10. M. van Kampen, C. Jozsa, J. T. Kohlhepp, P. LeClair, L. Lagae, W. J. M. de Jonge, and B. Koopmans, All-Optical Probe of Coherent Spin Waves, Phys. Rev. Lett. **88**, 227201 (2002).
11. A. De, A. Lentfert, L. Scheuer, B. Stadtmüller, G. Von Freymann, M. Aeschlimann, and P. Pirro, Coherent and incoherent magnons induced by strong ultrafast demagnetization in thin permalloy films, Phys. Rev. B **109**, 024422 (2024).
12. Y. Au, M. Dvornik, T. Davison, E. Ahmad, P. S. Keatley, A. Vansteenkiste, B. Van Waeyenberge, and V. V. Kruglyak, Direct Excitation of Propagating Spin Waves by Focused Ultrashort Optical Pulses, Phys. Rev. Lett. **110**, 097201 (2013).
13. R. A. Leenders, D. Afanasiev, A. V. Kimel, and R. V. Mikhaylovskiy, Canted spin order as a platform for ultrafast conversion of magnons, Nature **630**, 335 (2024).
14. Q. Remy and S. Mangin, From ultrafast demagnetization to ultrafast Spintronics: a 30-year story, Journal of Magnetism and Magnetic Materials **655**, 174344 (2026).
15. B. H. Rimmler, B. Pal, and S. S. P. Parkin, Non-collinear antiferromagnetic spintronics, Nat Rev Mater **10**, 109 (2024).
16. J. A. C. Incorvia et al., Spintronics for achieving system-level energy-efficient logic, Nat Rev Electr Eng **1**, 700 (2024).
17. J. Han, R. Cheng, L. Liu, H. Ohno, and S. Fukami, Coherent antiferromagnetic spintronics, Nat. Mater. **22**, 684 (2023).
18. B. Dieny et al., Opportunities and challenges for spintronics in the microelectronics industry, Nat Electron **3**, 446 (2020).
19. B. T. Fichera et al., Light-Induced Reorientation Transition in an Antiferromagnetic Semiconductor, Phys. Rev. X **15**, 011044 (2025).
20. S. Parchenko et al., Transient Non-Collinear Magnetic State for All-Optical Magnetization Switching, Advanced Science **10**, 2302550 (2023).
21. B. Ilyas et al., Terahertz field-induced metastable magnetization near criticality in FePS3, Nature **636**, 609 (2024).
22. J. Zhang, X. Tan, M. Liu, S. W. Teitelbaum, K. W. Post, F. Jin, K. A. Nelson, D. N. Basov, W. Wu, and R. D. Averitt, Cooperative photoinduced metastable phase control in strained manganite films, Nature Mater **15**, 956 (2016).
23. X. Chen et al., Ultrafast demagnetization in ferromagnetic materials: Origins and progress, Physics Reports **1102**, 1 (2025).
24. S. Parchenko, D. Pecchio, R. Mondal, P. M. Oppeneer, and A. Scherz, Magnetization precession after non-collinear dual optical excitation, Journal of Applied Physics **135**, 173902 (2024).
25. S. Parchenko, M. Riepp, S. Marotzke, A. Åberg Larsson, V. Kapaklis, P. M. Oppeneer, and A. Scherz, Demagnetization dynamics after noncollinear dual optical excitation, Phys. Rev. B **110**, 054425 (2024).
26. S. Parchenko, A. Tsukamoto, P. M. Oppeneer, and A. Scherz, Magnetization switching in GdFeCo induced by dual optical excitation, Phys. Rev. B **110**, 174401 (2024).
27. S. Parchenko, P. M. Oppeneer, and A. Scherz, Anomalous Magnetization Dynamics After Dual Optical Excitation, Advanced Physics Research **5**, e00057 (2026).

28. P. Carrara et al., All-Optical Generation and Time-Resolved Polarimetry of Magnetoacoustic Resonances via Transient Grating Spectroscopy, Phys. Rev. Applied **18**, 044009 (2022).
29. M. Brioschi et al., Excitation of Spin Waves in Ferrimagnetic Alloy via Optical Transient Grating Spectroscopy, Advanced Photonics Research **7**, e202500233 (2026).
30. G. Cao, S. Jiang, J. Åkerman, and J. Weissenrieder, Femtosecond laser driven precessing magnetic gratings, Nanoscale **13**, 3746 (2021).
31. L. Foglia et al., Nanoscale polarization transient gratings, Nat Commun **15**, 10742 (2024).
32. F. Caruso et al., The 2025 roadmap to ultrafast dynamics: frontiers of theoretical and computational modeling, J. Phys. Mater. **9**, 012501 (2026).
33. C. D. Stanciu, A. V. Kimel, F. Hansteen, A. Tsukamoto, A. Itoh, A. Kirilyuk, and Th. Rasing, Ultrafast spin dynamics across compensation points in ferrimagnetic GdFeCo : The role of angular momentum compensation, Phys. Rev. B **73**, 220402 (2006).
34. A. Tsukamoto, T. Sato, S. Toriumi, and A. Itoh, Precessional switching by ultrashort pulse laser: Beyond room temperature ferromagnetic resonance limit, Journal of Applied Physics **109**, 07D302 (2011).
35. V. Kambersky, On ferromagnetic resonance damping in metals, Czech. J. Phys. **26**, 1366 (1976).
36. V. Kambersky, On the Landau-Lifshitz relaxation in ferromagnetic metals, Can. J. Phys. 48, 2906 (1970).
37. T. L. Gilbert, A phenomenological theory of damping in ferromagnetic materials, IEEE Trans. Magn. 40, 3443 (2004).
38. K. Gilmore, Y. U. Idzerda, and M. D. Stiles, Spin-orbit precession damping in transition metal ferromagnets, Journal of Applied Physics **103**, 07D303 (2008).
39. K. Gilmore, M. D. Stiles, J. Seib, D. Steiauf, and M. Fähnle, Anisotropic damping of the magnetization dynamics in Ni, Co, and Fe, Phys. Rev. B **81**, 174414 (2010).
40. M. Fähnle and C. Illg, Electron theory of fast and ultrafast dissipative magnetization dynamics, J. Phys.: Condens. Matter **23**, 493201 (2011).
41. W. Zhang, W. He, X.-Q. Zhang, Z.-H. Cheng, J. Teng, and M. Fähnle, Unifying ultrafast demagnetization and intrinsic Gilbert damping in Co/Ni bilayers with electronic relaxation near the Fermi surface, Phys. Rev. B **96**, 220415 (2017).
42. F. Reyes-Osorio and B. K. Nikolić, Gilbert damping in metallic ferromagnets from Schwinger-Keldysh field theory: Intrinsically nonlocal, nonuniform, and made anisotropic by spin-orbit coupling, Phys. Rev. B **109**, 024413 (2024).